\documentclass[twocolumn,tighten]{aastex63}
\usepackage{amsmath}
\usepackage{amstext}

\usepackage{xcolor}

\shorttitle{Simulating the Eclipsing Binary Yield of the Vera Rubin Observatory}
\shortauthors{Geller et al.}

\newcommand{\baseline}{\texttt{baseline}}
\newcommand{\colossus}{\texttt{colossus}}

\begin{document}

\title{Simulating Eclipsing Binary Yields of the Rubin Observatory in the Galactic Field and Star Clusters}


\author[0000-0002-3881-9332]{Aaron M.\ Geller}
\affiliation{Center for Interdisciplinary Exploration and Research in Astrophysics (CIERA) and Department of Physics and Astronomy, Northwestern University, 1800 Sherman Ave., Evanston, IL 60201, USA}
\affiliation{Adler Planetarium, Department of Astronomy, 1300 S. Lake Shore Drive, Chicago, IL 60605, USA}
\email{a-geller@northwestern.edu}

\author[0000-0002-5283-933X]{Ava Polzin}
\affiliation{Center for Interdisciplinary Exploration and Research in Astrophysics (CIERA) and Department of Physics and Astronomy, Northwestern University, 1800 Sherman Ave., Evanston, IL 60201, USA}
\affiliation{Yale Department of Astronomy, Steinbach Hall, 52 Hillhouse Avenue, New Haven, CT 06511, USA}

\author{Andrew Bowen}
\affiliation{Center for Interdisciplinary Exploration and Research in Astrophysics (CIERA) and Department of Physics and Astronomy, Northwestern University, 1800 Sherman Ave., Evanston, IL 60201, USA}

\author[0000-0001-9515-478X]{Adam A.\ Miller}
\affiliation{Center for Interdisciplinary Exploration and Research in Astrophysics (CIERA) and Department of Physics and Astronomy, Northwestern University, 1800 Sherman Ave., Evanston, IL 60201, USA}
\affiliation{Adler Planetarium, Department of Astronomy, 1300 S. Lake Shore Drive, Chicago, IL 60605, USA}

\begin{abstract}
We present a study of the detection and recovery efficiency of the Rubin Observatory for detached eclipsing binaries (EBs) in the galactic field, globular clusters (GCs) and open clusters (OCs), with a focus on comparing two proposed observing strategies: a standard cadence (\baseline), and a cadence which samples the galactic plane more evenly (\colossus).  We generate realistic input binary populations in all observing fields of the Rubin Observatory,  simulate the expected observations in each filter, and attempt to characterize the EBs using these simulated observations.  In our models, we predict the \baseline\ cadence will enable the Rubin Observatory to observe about three million EBs; our technique could recover and characterize nearly one million of these in the field and thousands within star clusters.  If the \colossus\ cadence is used, the number of recovered EBs would increase by an overall factor of about 1.7 in the field and in globular clusters, and a factor of about three in open clusters.  Including semi-detached and contact systems could increase the number of recovered EBs by an additional factor of about 2.5 to 3. Regardless of the cadence, observations from the Rubin Observatory could reveal statistically significant physical distinctions between the distributions of EB orbital elements between the field, GCs and OCs.  Simulations such as these can be used to bias correct the sample of Rubin Observatory EBs to study the intrinsic properties of the full binary populations in the field and star clusters.

\end{abstract}


\section{Introduction}\label{s:intro}


The Vera C.\ Rubin Observatory (Rubin Observatory, formerly the Large Synoptic Survey Telescope) will allow for the most comprehensive survey of eclipsing binaries (EBs) in the southern sky to date when it begins operating in the early 2020s \citep{lsst2009}.  In particular the Rubin Observatory will undertake a 10-year survey of the southern sky, known as the Legacy Survey of Space and Time (LSST), which will return a tidal wave of light curve data.  \citet{prsa2011} predict that the  Rubin Observatory should observe approximately 24 million EBs with a signal-to-noise $>$10, and be able to characterise the orbits of about 30\% of these. When combined with radial-velocity measurements, e.g., from Gaia and/or SDSS/APOGEE, a subsample of these EBs will provide a large database of stars with measured radii, surface temperatures, luminosities, and dynamically-measured masses in our Galaxy, that can be used to test and constrain stellar evolution models, and underpin many fundamental studies in astrophysics \citep[e.g. see][]{andersen1991, torres2002, torres2010, thompson2010, prsa2011, brogaard2012, sandquist2013, jeffries2013, pietrzynski2013, brewer2016}.  In this paper, we examine the predicted recovery rate for detached EBs with the Rubin Observatory in the galactic field and in star clusters.

There have been a number of papers that explore the expected EB yield of the Rubin Observatory in the field, using different approaches.  Initial results can be found in \citet{lsst2009}. \citet{prsa2011} refined these results by employing a sample of 10,000 binaries with periods between 0.5 and 1,000 days and utilizing a neural network to analyze output results. \citet{wells2017} later explored a sample of $\sim$3000 already known and characterized EBs from the Kepler survey, which primarily have periods less than 10 days.

\begin{figure}[!t]
	\plotone{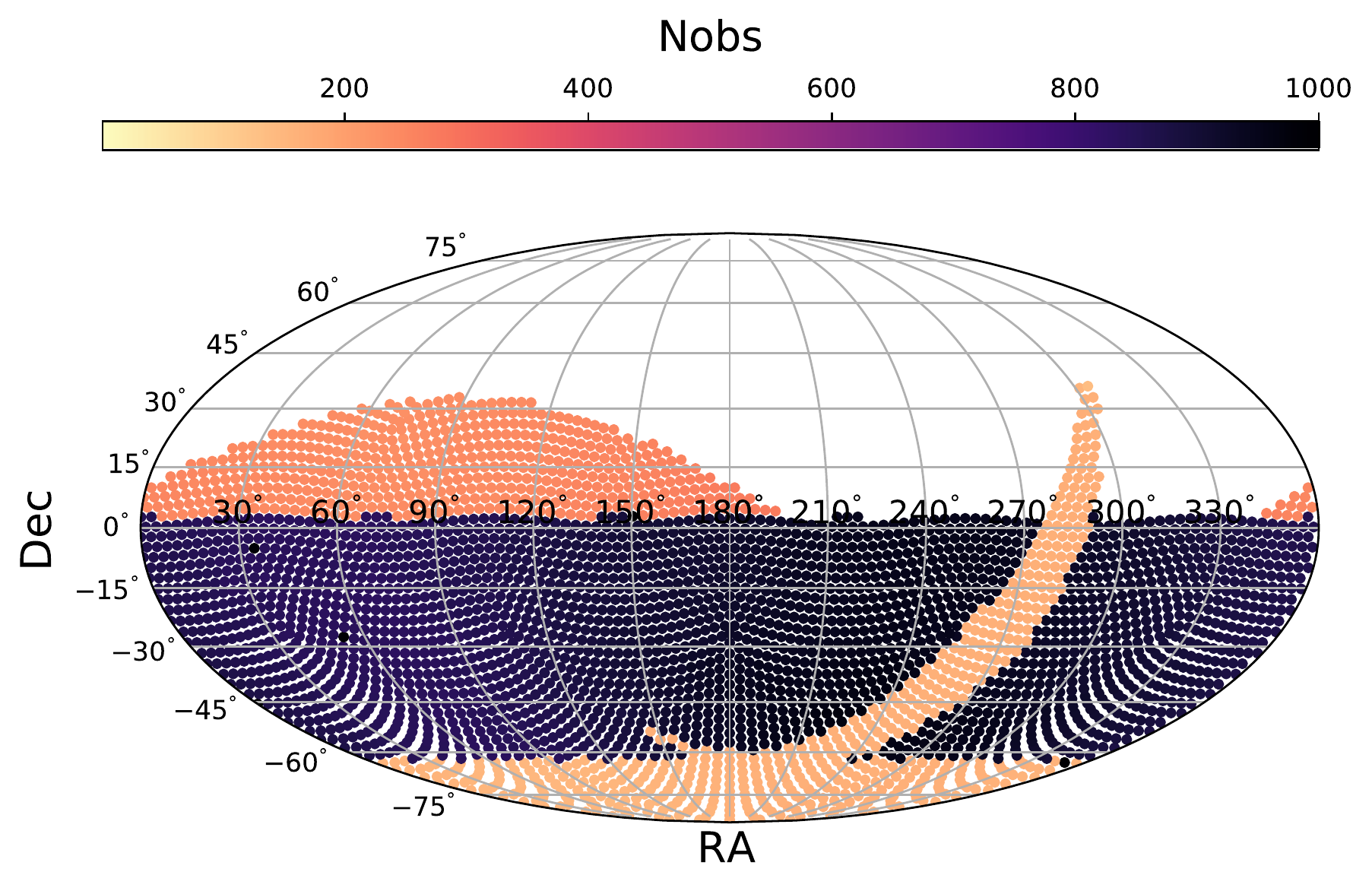}\vspace{-0.65em}
	\plotone{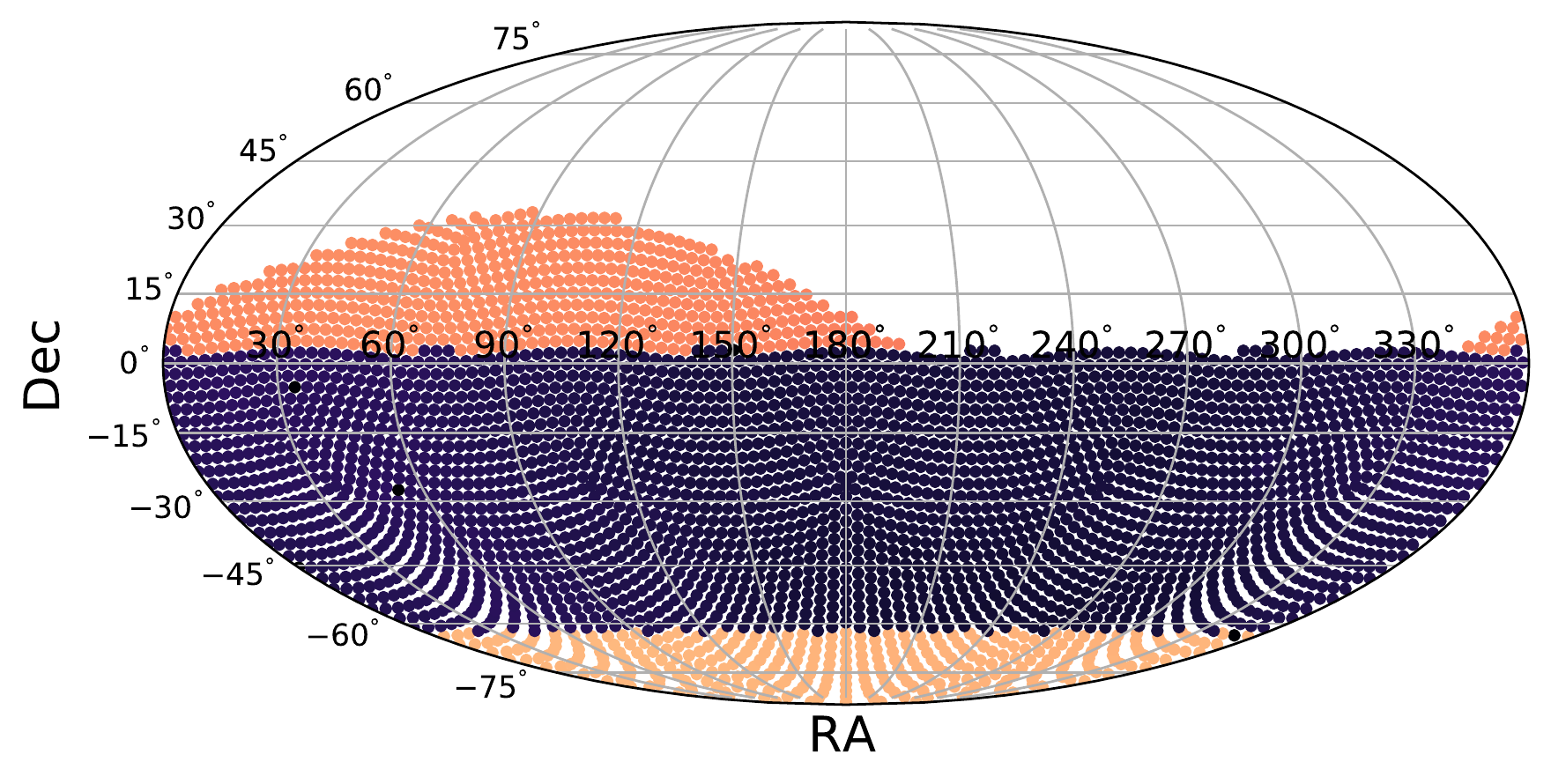}\vspace{1.7em}
	\plotone{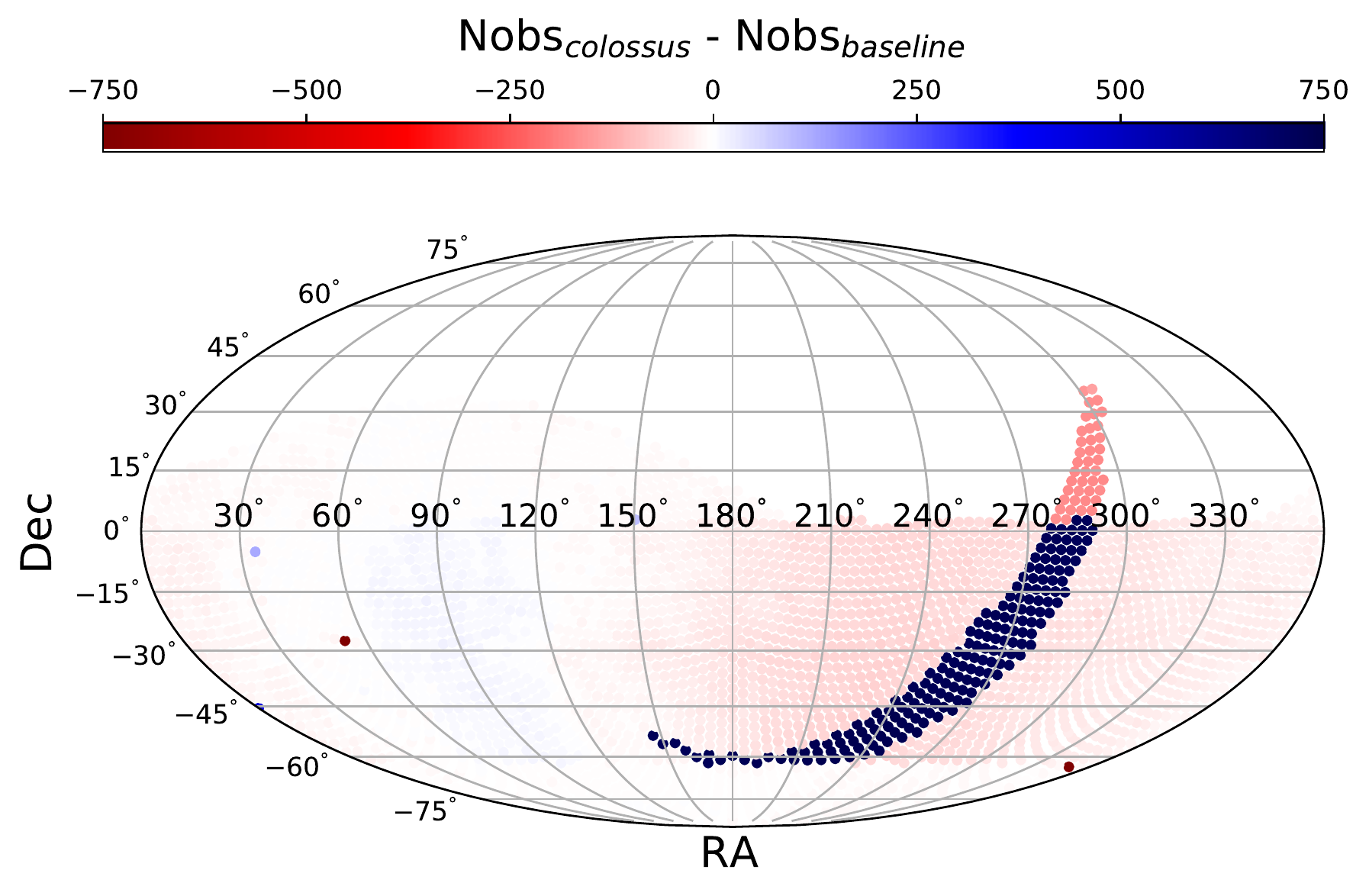}\vspace{1.7em}
	\plotone{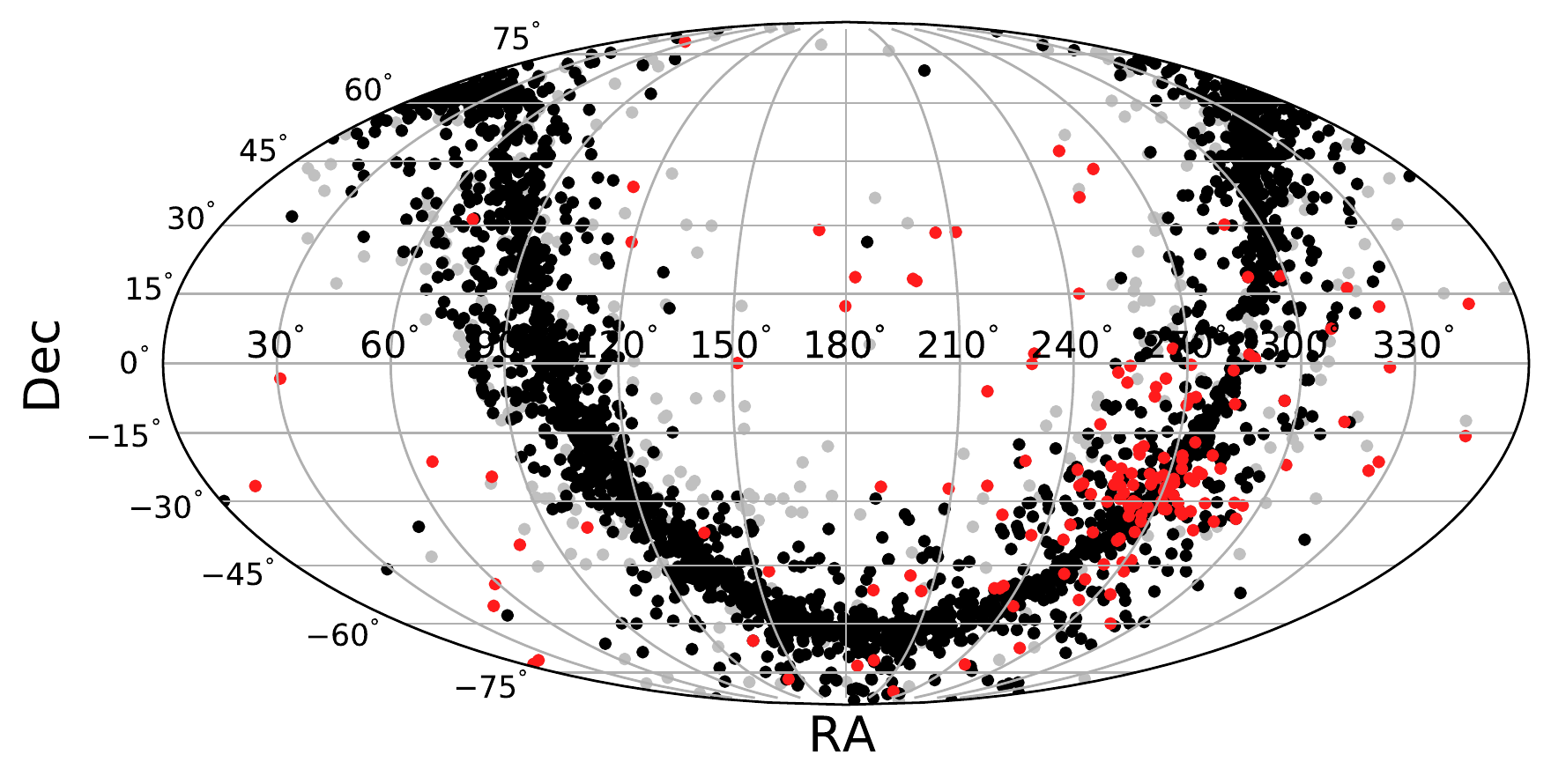}
	\caption{Sky distributions of the proposed Rubin Observatory \baseline\ (top) and \colossus\ (second from top) \texttt{OpSim} cadences,  difference in number of total observations (over all filters) between these cadences (third from top), and star clusters (bottom), all in the Mollweide projection.  The top two panels share the same colorbar showing the total number of observations expected over the LSST duration across all filters.  In the bottom panel, red points show GCs, black points show the OCs that are included in our study, and gray points show the low-mass OCs that are excluded from our study which are not expected statistically to have any EBs. 
    \label{f:OpSimClusterDist}}
\end{figure}

Like \citet{prsa2011}, in this paper we generate observations for individual binaries using \texttt{OpSim} \citep{delgado2014} -- the Rubin Observatory's own observation simulator -- to generate observations with the appropriate cadence and limiting magnitudes over the full observing area of the Rubin Observatory.  We  compare the results derived from two specific cadences: \texttt{baseline2018a} (hereafter, \baseline), the default cadence at the time of writing this paper, and \texttt{colossus\_2664} (hereafter \colossus), a proposed cadence that samples the galactic plane more evenly.  The total number of observations in each observing field of the Rubin Observatory are plotted in the top two panels of Figure~\ref{f:OpSimClusterDist}.  In this investigation, we simulate a far larger number of EBs than in previous papers, and use the (simulated) photometry from all Rubin Observatory filters in attempts to fully study the recovery statistics across the entire Rubin Observatory field of view. 

Additionally, we focus on the EBs expected to be observed by the Rubin Observatory in the Milky Way's globular clusters (GCs) and open clusters (OCs);  their distribution on the sky is shown in the bottom panel of Figure~\ref{f:OpSimClusterDist}, and the predicted LSST observing cadence is shown in the upper panels of the same figure.  From a comparison of the panels in Figure~\ref{f:OpSimClusterDist}, it is clear that a focus on the galactic plane may be important for observing star clusters.  EBs in star clusters are particularly valuable because we can use measurements of the many cluster member stars to determine very precise distance, metallicity, reddening and age measurements for the EB.  In turn these cluster EBs can be used to precisely constrain stellar evolution models \citep[e.g.][]{thompson2010, brogaard2012, sandquist2013, jeffries2013, brewer2016}.  Of additional interest, $N$-body models predict that stellar dynamics can shape the orbital elements and masses of binaries in star clusters through close encounters \citep{fregeau2004, hurley2005, chatterjee2010, geller2013, geller2013b, geller2015b, geller2019}.  Comparing the distributions of orbital elements and masses between binaries within the field to those in clusters can empirically probe the impacts of stellar dynamics and test the predictions of $N$-body models.  Large samples of binaries are needed for these comparisons, and the Rubin Observatory will be an important contributor.

The paper structure is as follows.  In Section~\ref{s:simObs} we explain our method to simulate EBs and observe them with the Rubin Observatory.  In Section~\ref{s:EByield} we analyze these results and make predictions for the EB yield of the Rubin Observatory with both cadences and in both the galactic field and in star clusters.  In Sections~\ref{s:discuss}~and~\ref{s:conclusions} we discuss these results and provide a summary and our conclusions.

\section{Simulating Observations of our Galaxy with the Rubin Observatory}\label{s:simObs}

Our aim is to simulate statistically the EB population of the Milky Way, including star clusters, and focusing on the non-degenerate objects.  As a high-level overview of our method: We generate a large sample of all binaries, both eclipsing and non-eclipsing, so that we can study how the predicted EBs recovered with the Rubin Observatory may map back to the instrinsic population.  We model the \baseline\ and \colossus\ \texttt{OpSim} cadences separately (see Figure~\ref{f:OpSimClusterDist}).  In each \texttt{OpSim} observing field (i.e., each field expected to be observed by the Rubin Observatory during the LSST survey), we sample from a \texttt{TRILEGAL} \citep{girardi2012} galactic model for the field of our galaxy, and if a star cluster resides in that \texttt{OpSim} field, we use \texttt{COSMIC} to evolve binaries appropriate for that star cluster.  If a given binary has an orbital period less than the LSST survey duration (3650 days), we include this binary in our \textit{``all"} sample.  If a given binary passes this period criterion and is oriented such that it eclipses, we use \citet{castelli2003} ATLAS9 model stellar atmospheres to derive the magnitudes in all of the Rubin Observatory filter bandpasses.  We then use the rapid light curve generator \texttt{ellc} \citep{maxted2016} to produce light curves for the binary in each of the Rubin Observatory filters and with the cadence for the specific observing field defined by \texttt{OpSim}.  If there is an eclipse observed with a $\geq3\sigma$ depth (where photometric uncertainties are derived from \citealt{ivezic2019}) and the binary has magnitudes within the limits of the Rubin Observatory,
we consider this binary \textit{``observable"} (and we will use this nomenclature throughout the paper).  For these \textit{observable} binaries, we use the \texttt{gatspy} code to attempt to recover the eclipse period.  If we can indeed recover the period with \texttt{gatspy} we consider these binaries \textit{``recoverable"} (and, likewise, we will use this nomenclature throughout the paper).  We run this procedure for each \texttt{OpSim} field in an MPI parallelized code on Northwestern's Quest High Performance Computer cluster.  We describe our procedure in greater detail in the following section.

\subsection{Sampling Eclipsing Binaries from the Galactic Field}\label{ss:sampleField}

For each \texttt{OpSim} field, we download a \texttt{TRILEGAL} v1.6 model using the \texttt{vespa} python wrapper \citep{morton2012}, with some minor code edits to better serve our purposes.  The \texttt{TRILEGAL} model returns a set of single stars appropriate for the given location in the galaxy, including their stellar parameters (e.g., mass, $\log(g)$, luminosity, temperature, metallicity, reddening, etc.).  We limit our sample to include only those stars with magnitudes $r<25$.  For the given observing field, we then create 40,000 binaries drawing from this \texttt{TRILEGAL} sample.  We choose the number 40,000 in order to ensure that each \texttt{OpSim} field is sampled evenly, and would result in a statistically robust estimate of the parameter distributions of the resulting EBs, assuming that of order 1\% of these binaries are detached and eclipsing, while also producing a computationally feasible number of binaries to analyze over all \texttt{OpSim} fields and our multiple different tests (as described below).   

We select stars randomly from the \texttt{TRILEGAL} model, and for each we create a binary as follows.  We sample a period, eccentricity and mass ratio from distributions consistent with observed binaries in the galactic field \citep{raghavan2010, geller2012, duchene2013, moe2017}.  

Specifically, we draw periods from a log-normal distribution with $\sigma = 2.28$ and a mean of 5.03 (in $\log$ days); we note that this empirical result is for solar-type stars and may not be fully appropriate for all stellar masses, but it is a decent approximation, and indeed most of our \textit{recoverable} EBs are near solar mass (see Figure~\ref{f:recdists}).  True field binaries have orbital periods extending beyond $10^{10}$ days; we limit our periods to be $<$3650 days so as not to sample binaries with periods longer than the expected observing duration of the LSST.  (We then account for this difference in our statistical analysis later on.)  This period limit will result in most EBs in our model undergoing at least one primary and one secondary eclipse during the LSST duration (except for some highly eccentric EBs), which is the minimum we would require to consider characterising the orbit. (Note that this still may not be sufficient to characterize the orbit; this criterion is simply meant to exclude EBs with periods we would not be able to recover with our method, in order to save on computation time.)

We sample eccentricities from a uniform distribution, but impose circular orbits for periods $<10$ days, due to tides (also consistent with observed field binaries).  \citet{moe2017} find some evidence for larger eccentricities in longer-period (10-500 day) binaries and also in binaries with higher-mass primaries; however in both cases the distributions are consistent with a uniform distribution to within $\sim$2$\sigma$.  We therefore choose to draw eccentricities from a uniform distribution for all binaries not effected by tides, and note that for non-circular binaries in the field our resulting recoverability statistics do not depend strongly on eccentricity (see top panel in Figure~\ref{f:recdists}).

We choose random orbital angles. Specifically, for each binary we draw the inclination randomly on the surface of the sphere, the longitude of periastron randomly between 0 and 2$\pi$, and we also choose a random date of the first mid-eclipse for star 1 by star 2 (within one orbital period of the start of the LSST survey).

To select a companion mass for a given binary, we first draw a random mass ratio from a uniform distribution, and then find the star in the \texttt{TRILEGAL} model with the nearest mass (which in practice limits our stellar masses to be $\gtrsim 0.1 M_\odot$).   This way we retain the expected stellar parameters predicted from \texttt{TRILEGAL} while also attempting to satisfy the empirically expected mass-ratio distribution.  Once this procedure is completed, the resulting mass-ratio distribution departs slightly from a uniform distribution, and especially for low-mass stars (as was studied in detail in \citealt{kouwenhoven2009}).  Nonetheless, the overall mass distribution is very close to uniform (see Figure~\ref{f:recdists}).

Finally, each binary also retains the distance, $A_V$, and metallicity for the primary star from the \texttt{TRILEGAL} model.

\subsection{Sampling Eclipsing Binaries from Galactic Star Clusters}
\label{ss:sampleClusters}

We use the \citet{harris2010} GC catalog and compile our own catalog of OCs, by combining data from the Milky Way Star Clusters Catalog \citep[MWSC;][]{kharchenko2012, kharchenko2013, schmeja2014, scholz2015}, WEBDA \citep{paunzen2008, netopil2012}, Lynga Open Cluster Catalog \citep{lynga1995}, and catalogs from \citet{salaris2004}, \citet{vandenbergh2006}, \citet{piskunov2008} and \citet{cantatgaudin2018}.  In total we compile a sample of 157 GCs and 3353 OCs\footnote{\footnotesize Interested readers can find the python scripts used to compile this OC catalog along with the resulting catalog here : \url{https://github.com/ageller/compileOCs}}.  For each of these clusters, we require values for the RA, Dec., distance, metallicity, age, total mass, half-mass radius and velocity dispersion.  

For the GCs, much of these data already exists in the \citet{harris2010} catalog.  We use the \citet{marinfranch2009} result to add ages for all the clusters.  We also estimate masses for the clusters assuming a mass-to-light ratio of 2.  When not already available in the \citet{harris2010} catalog, we calculate a central velocity dispersion using equations for a \citet{plummer1911} model.  Finally for the few clusters that do not have one (or more) of the required parameters, we simply take the mean values from the respective distributions.  (This, of course, is not ideal, but will result in distributions roughly consistent with those observed while allowing us to sample binaries from each cluster.)

Open clusters do not have such a well maintained catalog as the GCs, and therefore we have compiled together many catalogs (see above) in hopes of producing as complete a sample as possible.  We made every attempt to remove duplicate clusters coming from different catalogs (which might be named differently, and also might have slightly offset positions).  When there are multiple values for a given cluster from different catalogs for a particular parameter that we require (e.g., distance, age, etc.), we simply take the average of the available values.  As for the GCs, if a value is not available, we take the mean from the distribution, and we calculate the central velocity dispersions assuming a \citet{plummer1911} model.  We also only include clusters in our sample where we might expect to observe at least one eclipsing binary (again assuming $\sim$1\% of binaries are detached and eclipsing). This results in a sample of 2585 (out of the total of 3353) OCs that we run through our analysis pipeline.

In order to generate binaries from star clusters for our Rubin Observatory analysis, we use the \texttt{COSMIC} \citep{breivik2019} population synthesis code, which is an improved version of the rapid binary evolution code \texttt{BSE} \citep{hurley2002} with a python wrapper.  Again, we generate 40,000 binaries per cluster in order to ensure robust statistics while maintaining a computationally manageable sample.  We allow \texttt{COSMIC} to produce an initial sample of binaries using the \texttt{multidim} sampler, which samples from the \citet{moe2017} empirical field distributions, except that we impose a long-period limit on the initial orbital period distribution at the expected ``hard-soft boundary" of the given cluster or the LSST survey duration if that is shorter. At periods longer than the hard-soft boundary, binaries are expected to be quickly disrupted by gravitational encounters with other cluster members.  We estimate the hard-soft boundary using Equation 1 in \citet{geller2015}, which requires the cluster velocity dispersion, and we assume a binary with solar-mass stars encountering a 0.5 $M_\odot$ star (roughly the mean stellar mass in most star clusters).  We use the given cluster's metallicity and age, and allow \texttt{COSMIC} to evolve these binaries up to the present day, using the default parameters recommended by \texttt{COSMIC} for \texttt{BSE}. Finally, we sample the random orbital angles on the sphere and choose the random initial date of eclipse, in the same method as for the field.

\subsection{Determining if a Binary is Observable}

We are interested here in comparing the intrinsic distributions of binary parameters (e.g., the period, eccentricity and mass-ratio distributions) with those that are observable and those that are expected to be recovered with our analysis methods.  Therefore, we expect that a large majority of the binaries we generate will not be observable by the Rubin Observatory (e.g., because they are not eclipsing or because they are too faint).  Consequently, before we simulate observations from the Rubin Observatory, we first check if the binary is observable.  We have multiple criteria that we test in order, and only those that pass all tests have observations simulated and periods analyzed, in order to save on compute time. 

\begin{enumerate}
\item \textit{Eclipsing} :  We require a geometry that results in an eclipse, given the inclination, stellar radii, semi-major axis, eccentricity and argument of periastron.  We include any eclipse depth at this step, only requiring that the stellar radii begin to overlap.

\item \textit{Detectable Period} :  We only include binaries with periods within the total expected observing duration of the LSST (10 years).  

\item \textit{Detached} :  For our analysis here, we are only interested in detached binaries (and this is also a limitation of \texttt{ellc}).  Therefore we exclude any binaries with either star filling its Roche lobe, using the equation from \citet{eggleton1983}.

\item \textit{Observable Magnitude} : If a binary passes these above criteria, we then use the ATLAS9 stellar atmospheres \citep{castelli2003} to compute the magnitudes of the binary in all Rubin Observatory filters (as described in Section~\ref{ss:LSSTobs}). We only include a given binary in the observable sample if the (total out-of-eclipse) $r$ magnitude of the binary is between 15.8 and 25 \citep{lsst2009}\footnote{\footnotesize Using, the $r$ magnitude for this limit will artificially exclude some very low-mass stars, given their relatively large colors in redder filters with respect to $r$; we investigated for this effect in a random observing field and found that if we were to instead limit by the $z$ or $y$ filters we would include a few percent more M-dwarf stars at the expense of removing a somewhat larger amount of higher-mass stars.  For studies particularly interested in M-dwarfs, a more carefully crafted criteria for observability would be warranted (though that is beyond the scope of this paper).}.

\item \textit{Sufficient Eclipse Depth} : If a binary passes all the criteria above, we then calculate the light curve using \texttt{ellc} (described in Section~\ref{ss:LSSTobs}), and sample it on the cadence from \texttt{OpSim}.  We use the method of \citet{ivezic2019} to include the expected random errors on the observations.  We then calculate the maximum depth observed for the eclipse in the $r$ band, and divide this by the expected uncertainty on the magnitude in that observation, as an estimate of the signal-to-noise of the eclipse.  If this signal-to-noise value is $\geq3$, we run this binary through \texttt{gatspy} to attempt at recovering the eclipse period. 

\end{enumerate}

\subsection{Simulating Observations from the Rubin Observatory}\label{ss:LSSTobs}

For each star in a binary, we use the $\log(g)$, metallicity, and effective temperature to define an ATLAS9 model stellar atmosphere \citep{castelli2003} .  We then use the Rubin Observatory filter bandpasses \citep{lsst2009}, along with the stellar radius, distance and luminosity to calculate the flux within each of the Rubin Observatory filters for each member of the binary.  For each filter, we simply add the fluxes of both stars together to get the combined flux of the binary (out of eclipse) in that filter.  We use the \citet{fitzpatrick04} extinction model with $R_V=3.1$, and the $A_V$ value for the given binary, to account for reddening.  

For the binaries in the galactic field, we could potentially use the Rubin Observatory magnitudes provided in the \texttt{TRILEGAL} model instead; however we do not have that option for the cluster binaries.  Therefore, for consistency, we follow our procedure for all binaries.  As a check, we compared our derived magnitudes to those from the \texttt{TRILEGAL} model and find very close agreement. 

After deriving the combined flux for the binary out of eclipse, we use the \texttt{ellc} code \citep{maxted2016} to model the eclipse.  We use the limb-darkening law from \citet{claret2000}, and assume a spherical shape for all stars in \texttt{ellc}.  We use \texttt{ellc}'s \texttt{light\_3} input to account for background flux due to crowding (as explained in Section~\ref{ss:crowding}).  Observing dates are defined by the \texttt{OpSim} model for the given observing field\footnote{\footnotesize For star clusters we take the nearest \texttt{OpSim} field, and assume all observations of the cluster will have that cadence}; in this paper we test two models.  The \baseline\ model provides the default expected LSST cadence (at the time of writing), which limits observations near the galactic center. The \colossus\ model weights the galactic plane more equally (which may be particularly important for the star clusters).  See Figure~\ref{f:OpSimClusterDist} for the number of observations in each observing field of the LSST within each of these models and for the different in number of observations between these two cadences.  For comparison, the star clusters are plotted in the same projection in bottom panel of Figure~\ref{f:OpSimClusterDist}.

We multiply the out-of-eclipse fluxes by the light curve produced by \texttt{ellc}, in each respective filter, and convert these to magnitudes (accounting for reddening). We follow \citet{ivezic2019} to derive the random error, given each magnitude and the $m_5$ value from \texttt{OpSim}.  The result is a realistic light curve with uncertainties in each of the Rubin Observatory filters and at the cadence predicted by the respective \texttt{OpSim} model.  We show an example of a light curve generated through this method in Figure~\ref{f:LCcrowding}.


\subsection{Crowding}\label{ss:crowding}

\begin{figure*}[!t]
    \centering
    \includegraphics[height=0.75\textheight]{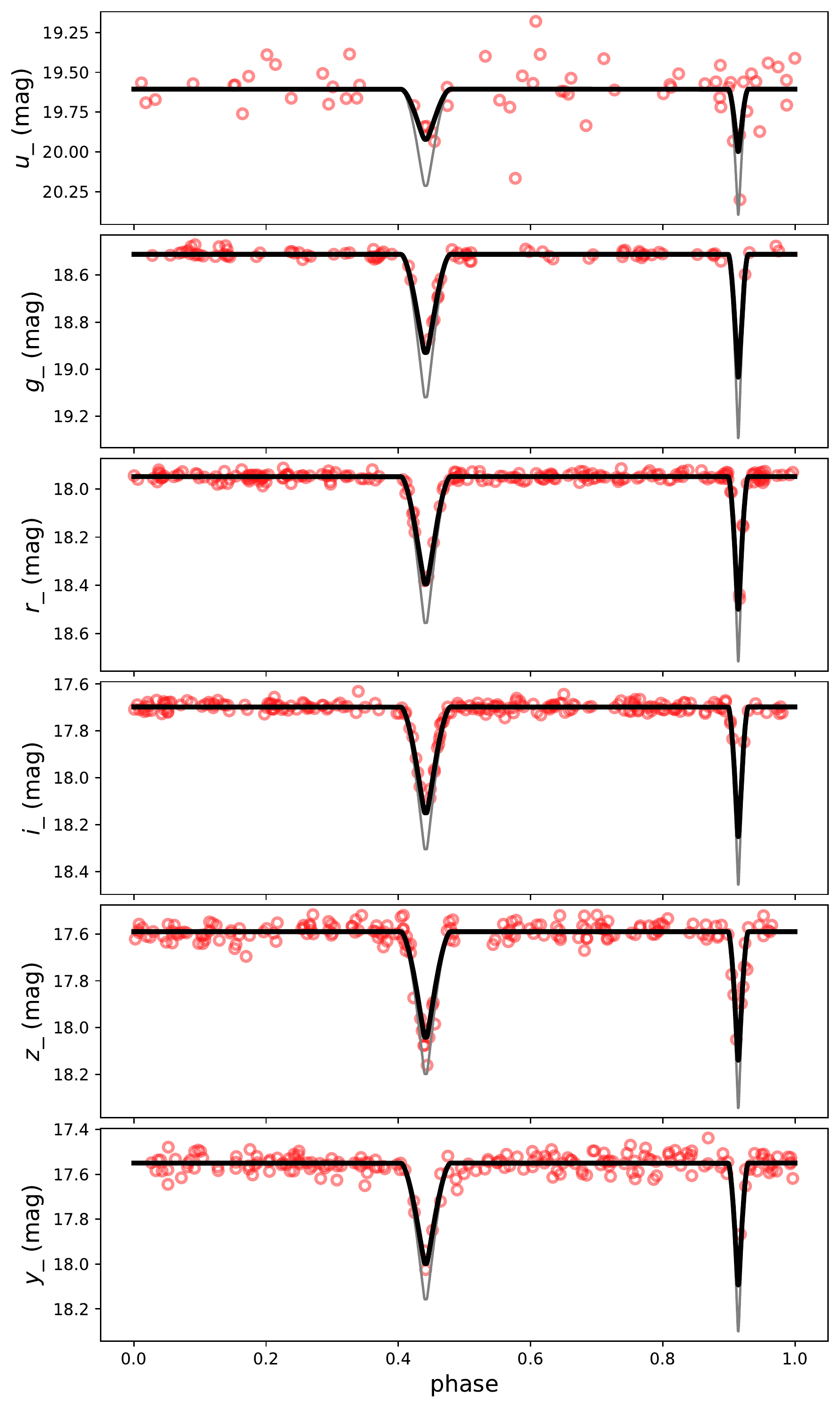}
    \includegraphics[height=0.75\textheight]{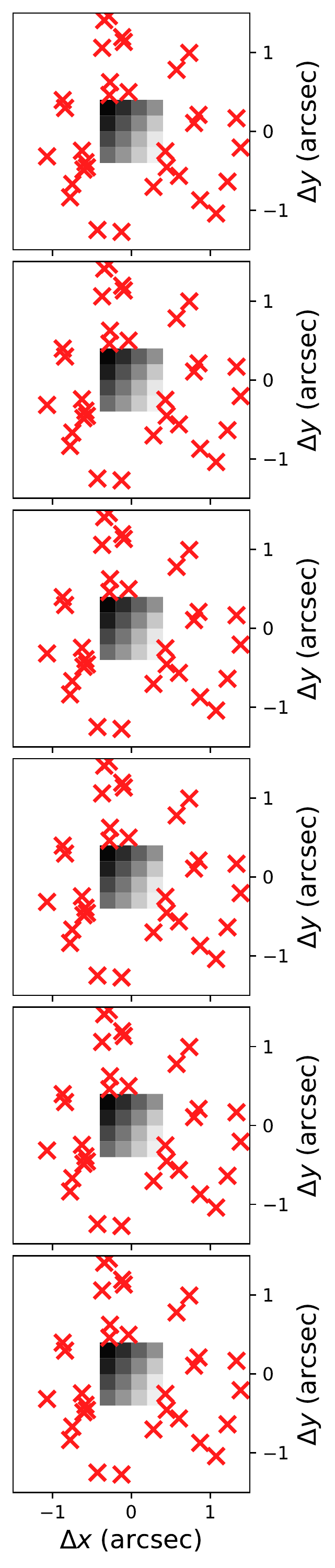}
	\caption{Simulated light curve for a random binary near the center of the GC M10, with crowding applied.  In the left panel, thin gray lines show the light curve without crowding, thick black lines show the light curve with crowding applied, and red points show the light curve sampled on the observing dates from \baseline\, with random errors applied.  On the right, we show the pixel map used to calculate the flux contribution from nearby stars.  The red ``X''s show the locations of simulated stars.  The grayscale regions show the pixels that are considered in the crowding calculation (within one resolution element, defined by an assumed seeing of 0.5 arcsec, from the observed binary), with darker regions indicating more flux.  \label{f:LCcrowding}}
\end{figure*}

Though the Rubin Observatory will have small pixels ($\sim$0.2 arcseconds on the sky), and the seeing is expected to be $\sim$0.5 arcseconds, there will still be issues with crowding in certain fields, and especially when observing the GCs and toward the galactic center.  We account for crowding in our analysis for both clusters and the field, where necessary, using the following procedure.  We focus only on single stars as contaminants here, and do not include additional simulated foreground and/or background EBs in the crowding calculation. 

For all binaries we include the crowding due to the galactic field.  Here we use the number of stars in the particular \texttt{OpSim} field from the \texttt{TRILEGAL} model divided by  the Rubin Observatory field-of-view to get a ``surface density" of stars.  We then use a 0.5 arcsecond seeing value to calculate the number of expected stars per resolution element.  (In principle we could use a more sophisticated measure of the resolution element that depends on the \texttt{OpSim} field and expected observing conditions, though this will not likely change our results significantly.)  We then use this value to estimate the number of stars expected within a distance of three times the seeing.  This value is chosen to allow very bright stars to still be included even if they are separated by a few resolution elements from the binary of interest.  If $\geq$1 stars are expected within this area on the sky, we draw that number of single stars randomly from the \texttt{TRILEGAL} model for that \texttt{OpSim} field,  distribute them uniformly over that circular area, and derive their respective fluxes as described in Section~\ref{ss:LSSTobs}.

For binaries within star clusters, we also check for crowding from other cluster stars.  To do so, we first draw a random position for the binary from a \citet{plummer1911} model, consistent with the observed cluster half-mass radius.  We then generate random positions for the expected number of stars in the cluster from the same \citet{plummer1911} model, and  select only those within a radius of three times the seeing from our binary of interest. If there are $\geq$1 stars expected in this area, we use \texttt{COSMIC} to evolve that number of single stars, drawn from a \citet{kroupa1993} IMF and with the cluster's metallicity, up to the cluster's age.  We follow the same procedure as above to calculate each star's expected flux in each of the Rubin Observatory filters.  

We then use the locations and fluxes of these singles stars (from both the galaxy and also the cluster, where relevant), with an assumed Gaussian point-spread-function having a width equal to the assumed seeing, to sum up the expected flux within all pixels inside of one seeing area.  This total flux is then used to define the ``\texttt{light\_3}" input to \texttt{ellc}, in order to account for crowding.  In general, fields near the galactic center, and many of the GCs, suffer from substantial crowding, while most EBs in other fields are not expected to be hindered by significant crowding.  We show the results from one such crowding calculation in Figure~\ref{f:LCcrowding}.



\section{Predicting the Rubin Observatory Eclipsing Binary Yield}\label{s:EByield}

\subsection{Period Analysis}



For observable binaries that have $\geq$3$\sigma$ eclipse depths, we use the \texttt{gatspy} code \citep{vanderplas2015} to attempt to recover the orbital period.   \texttt{gatspy} has the option to use a multiband periodogram, which extends the usual Lomb-Scargle period analysis (which traditionally can only accommodate observations in one filter) to include observations from multiple filters.  For our purposes we attempt to derive the period in each of the Rubin Observatory filters separately as well as in combination (and we discuss these results below).  We choose to use the ``fast" algorithm from \texttt{gatspy}, to make the analysis more computational tractable.  The fast algorithm does not allow the user to introduce additional harmonics into the period search (while their other algorithm does); we tested the other algorithm with various additional terms (``band" and ``base" in \texttt{gatspy}), and find a few percent improvement overall in the number of EBs we could recover, but not enough to warrant the much higher computational overhead it introduced for this project.   

In our analysis we check to see if the \texttt{gatspy} period is within 10\% of either the true full period, half the true period or twice the true period.  Those that pass this test are considered \textit{recoverable} in our analysis.  (We do not explicitly impose a minimum number of observed eclipses for an EB to be considered \textit{recoverable}, other than our initial limit of the orbital period.  We allow for identification of the half period and twice the period as these can sometimes produce the largest power in a periodogram analysis of real EBs, especially for circular EBs with near-equal brightness ratios.)

\begin{deluxetable*}{l r r r r}
    \label{t:summary}

    \tablecaption{Summary of Simulated Rubin Observatory Detached EBs}

    \tablehead{\colhead{Category} & \colhead{All} & \colhead{Obs.} & \colhead{Rec.} & \colhead{Rec./Obs.}}
    
    \startdata
        \multicolumn{5}{l}{\textbf{\baseline}}\\
        Field & $(1.00 \pm 0.05)\times10^9$   & $(2.86 \pm 0.05)\times10^6$  & $(9.34 \pm 0.16)\times10^5$ & 32.6\% $\pm$ 0.4\% \\
        GCs   & $(3.5 \pm 0.3)\times10^6$     & $(1.40 \pm 0.20)\times10^4$  & $(4.3 \pm 0.6)\times10^3$  & 30.4\% $\pm$ 1.2\%\\
        OCs   & $(1.40 \pm 0.21)\times10^5$   & $(5.7 \pm 1.3)\times10^2$    & $(1.3 \pm 0.3)\times10^2$  & 23.8\% $\pm$ 2.3\% \\
        &&&&\\
        \multicolumn{5}{l}{\textbf{\colossus}}\\
        Field & $(9.6 \pm 0.5)\times10^8$     & $(5.90 \pm 0.20)\times10^6$ & $(1.56\pm 0.05)\times10^6$ & 26.4\% $\pm$ 0.4\% \\
        GCs   & $(3.8 \pm 0.3)\times10^6$     & $(2.24 \pm 0.22)\times10^4$ & $(6.8 \pm 0.8)\times10^3$  & 30.2\% $\pm$ 1.1\%\\
        OCs   & $(1.38\pm 0.21)\times10^5$    & $(1.2 \pm 0.3)\times10^3$   & $(3.8 \pm 0.9)\times10^2$  & 31.2\% $\pm$ 2.3\% \\
        &&&&\\
        \multicolumn{5}{l}{\textbf{\colossus/\baseline}}\\
        Field & 0.96 $\pm$ 0.06 &   2.06 $\pm$ 0.08 &   1.67 $\pm$ 0.06 & \nodata \\
        GCs   & 1.08 $\pm$ 0.12 &   1.6 $\pm$ 0.3   &   1.6 $\pm$ 0.3 & \nodata \\
        OCs   & 1.01 $\pm$ 0.22 &   2.3 $\pm$ 0.8   &   3.0 $\pm$ 1.0 & \nodata \\
    \enddata

\end{deluxetable*}

\subsection{Normalization}

As described above, for each \texttt{OpSim} field in both the galactic field population and for star clusters, we choose to generate a statistical sample of 40,000 binaries.  Our first task in the analysis is to normalize this number so that each \texttt{OpSim} field is counted with the correct number of binaries that would be observed in either the galactic field or the respective star cluster.  In order to calculate the expected number of binaries in a given simulated sample, we first use the following equation.
\begin{equation}\label{e:norm}
    N = \left(\frac{N_\text{sample}}{N_\text{all}}\right)\left(N_\text{stars} \sum_i p(m_{1,i}) f_b(m_{1,i})\right) 
\end{equation}

Here $N$ is the expected real number of binaries in that \texttt{OpSim} field.  The first term on the right side of Equation~\ref{e:norm} is the fraction of stars we count in the given field;  $N_\text{sample}$ is the number we count within the given simulated binary sample of that field (e.g., in the \textit{recoverable} sample), and $N_\text{all}$ is the total number of simulated binaries in that field (e.g., 40,000).

The right-hand term in Equation~\ref{e:norm} is the normalization to the total number of binaries expected in the given field.  Here, $N_\text{stars}$ is the total number of stars expected in that field.  For our analysis of the galactic field, we take $N_\text{stars}$ as the total number of stars in the \texttt{TRILEGAL} model for that location on sky.  For a given cluster, we use the observed (or estimated) number of stars in the cluster for $N_\text{stars}$.  We define the primary (most massive star) mass of a given binary as $m_1$.  In this formula, we bin the primary masses in the sample in bins of 0.1 $M_\odot$, and $p(m_{1,i})$ is the fraction of primary stars within the given mass bin in our sample. 
The summation is included to weight the sample by the appropriate binary frequency expected for this primary mass, $f_b(m_{1,i})$.  We estimate this binary frequency by fitting a one-dimensional power-law to the results from \citet{raghavan2010} for the binary frequency as a function of mass, and find the following form:
\begin{equation}
    f_b(m_1) = 0.487 \alpha \left(\frac{m_1}{1.219}\right)^{0.184}
\end{equation}
Finally, $\alpha$ is the fraction of binaries within the full log-normal period distribution that have periods less than the imposed period cutoff (as described in Sections~\ref{ss:sampleField}~and~\ref{ss:sampleClusters}).  Note that for most star clusters, the full binary fraction is not known; our procedure here is similar to that of \citealt{geller2015}, where we assume a field-like binary population limited to the hard-soft boundary (or 3650 days, if shorter).  As a reference, $\alpha \sim 0.26$ for a period cutoff of 3650 days.

The resulting total numbers of binaries in each of our various simulations resulting from these calculations are listed in Table~\ref{t:summary}.  

\vspace{2em}
\subsection{Uncertainties} \label{ss:uncertainties}

In Table~\ref{t:summary}, we also provide uncertainty estimates for each of the summary values.  These result from the Poisson uncertainties on the values that we derive from our simulations and also the observational uncertainties on the binary frequency as a function of mass from \citet{raghavan2010}, which we use in Equation~\ref{e:norm}.   Specifically, for each \textit{OpSim} field, we ran 1000 Monte Carlo samples accounting for these uncertainties to derive a distribution of each respective value.  In each case, the resulting distribution is centered on the value derived directly from our simulation, and we provide the 1$\sigma$ width of this distribution as the uncertainty quoted in Table~\ref{t:summary}. 

\begin{figure*}[!t]
    \epsscale{1.15}
	\plotone{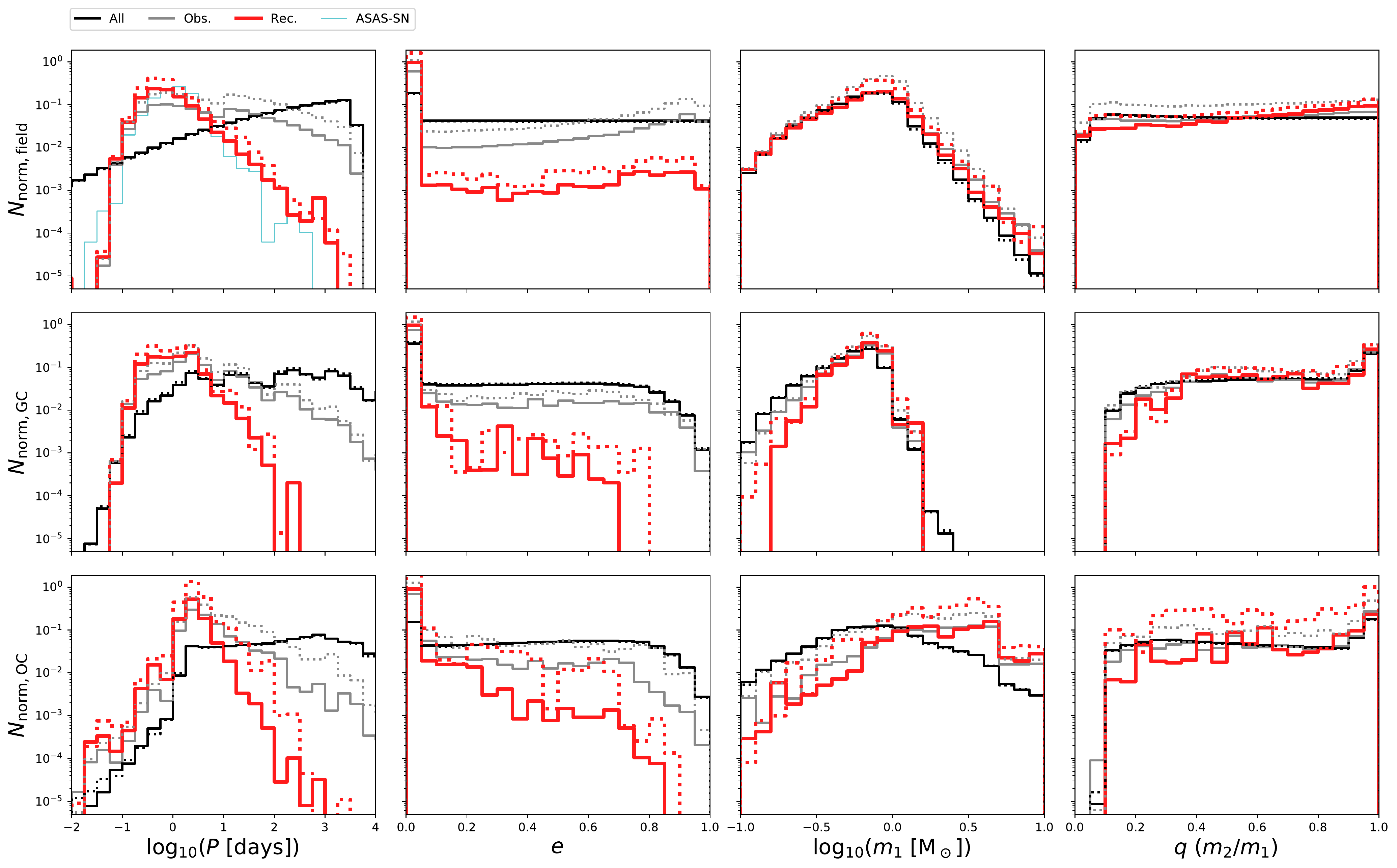}
	\caption{Distributions of orbital parameters and masses for the galactic field (top), GC (middle) and OC (bottom) samples of simulated Rubin Observatory EBs.  In each panel, the black lines show the full sample (``All"), the gray lines show the \textit{observable} sample (``Obs."), and the red lines show the \textit{recoverable} sample (``Rec.").  Solid lines show the results from our \baseline\ simulations, and dotted lines show results from our \colossus\ simulations.  Each histogram is normalized by dividing by the total number of binaries in the associated \baseline\ sample, respectively.  As a comparison, in the upper-left panel we also plot period distribution of the EA-type EBs identified in the ASAS-SN survey, normalized by the total number of EA-type EBs in ASAS-SN,  with the blue line.  From left to right we show the distributions of  orbital period, eccentricity, primary mass, and mass ratio.  \label{f:recdists}}
\end{figure*}

In addition to these quoted uncertainties, we also investigated the effects of possible differences in the intrinsic population as compared to our chosen initial conditions.  By inspecting Figures~\ref{f:recdists}~and~\ref{f:recratios}, one can see that the recovery rate is most sensitive to the period.  As discussed above, we draw periods from the log-normal distribution observed by \cite{raghavan2010} for field solar-type binaries.  If instead we set our initial period distribution to be log-uniform (a common distribution in theoretical binary analyses), we would expect an increase of a factor of $\sim$3 in the numbers of \textit{observable} and \textit{recoverable} binaries from the field model (either \baseline\ or \colossus).  To derive this factor we first constructed a uniform period distribution within our initial period limits and containing the same number of binaries as for our given model.  We then multiplied this distribution by the \textit{Rec.}/\textit{All} period distribution resulting from our model (seen in the red line in Figure~\ref{f:recratios}).  The number of binaries summed over this distribution represents the total number of \textit{recoverable} EBs expected from the initially uniform period distribution (which is $\sim$3 times larger than with the log-normal distribution).

Eccentricity is another important parameter that impacts our ability to characterise the orbit.  For the field, we choose an initially uniform eccentricity distribution that is then modified by tides.  Alternatively, some theoretical models choose a ``thermal" eccentricity distribution that rises toward larger eccentricities (though there is some doubt that this distribution does indeed occur in real binary populations, \citealt{geller2019}).  If instead we chose a thermal eccentriciy distribution for our initial conditions and modify it to account for tides as we do in our model, we estimate this would reduce the number of \textit{recoverable} binaries in the field by a factor of about 1.2.

\section{Investigating the Simulated Binaries in the Field and Star Clusters}

In the following sections we investigate each of our samples in more detail, beginning with the galactic field simulation.  Overall, our model predicts that there will be of order one billion binaries with appropriate magnitude limits and sky locations to be accessible to the Rubin Observatory, though only a small fraction of them will be eclipsing.  Using the \baseline\ cadence, our models predict that $\sim$3 million detached EBs will be \textit{observable} and nearly one million detached EBs would be \textit{recoverable} using our method.  Using the \colossus\ cadence, the Rubin Observatory would observe about two times more EBs and could potentially recover $\sim$1.7 times more EBs.

\begin{figure*}[!t]
    \epsscale{1.15}
	\plotone{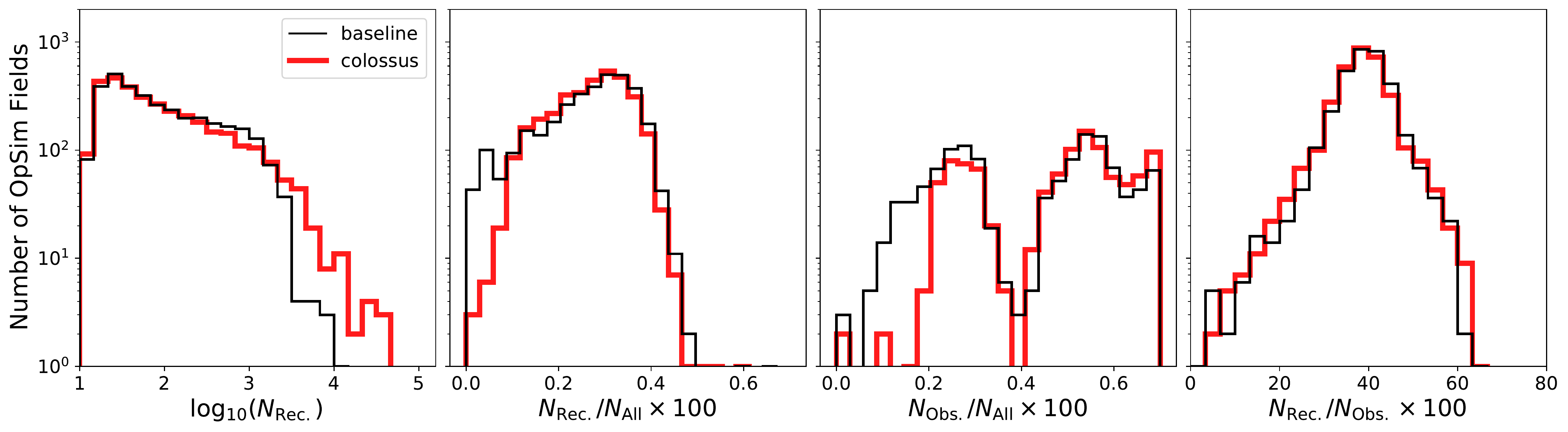}
	\caption{Histograms of recovery statistics for the galactic field model, comparing the \baseline\ cadence (black) and the \colossus\ cadence (red).  Each entry in a given histogram is one \texttt{OpSim} field. From left to right, we show histograms of the predicted number of \textit{recoverable} binaries ($N_{\rm Rec.}$), the ratio of the number of \textit{recoverable} binaries over all binaries ($N_{\rm Rec.}/N_{\rm All}$), the ratio of the number of \textit{observable} binaries over all binaries ($N_{\rm Obs.}/N_{\rm All}$), and the ratio of the number of \textit{recoverable} binaries over the number of \textit{observable} binaries ($N_{\rm Rec.}/N_{\rm Obs.}$).  \label{f:recHists}}
\end{figure*}

\begin{figure}[!t]
	\plotone{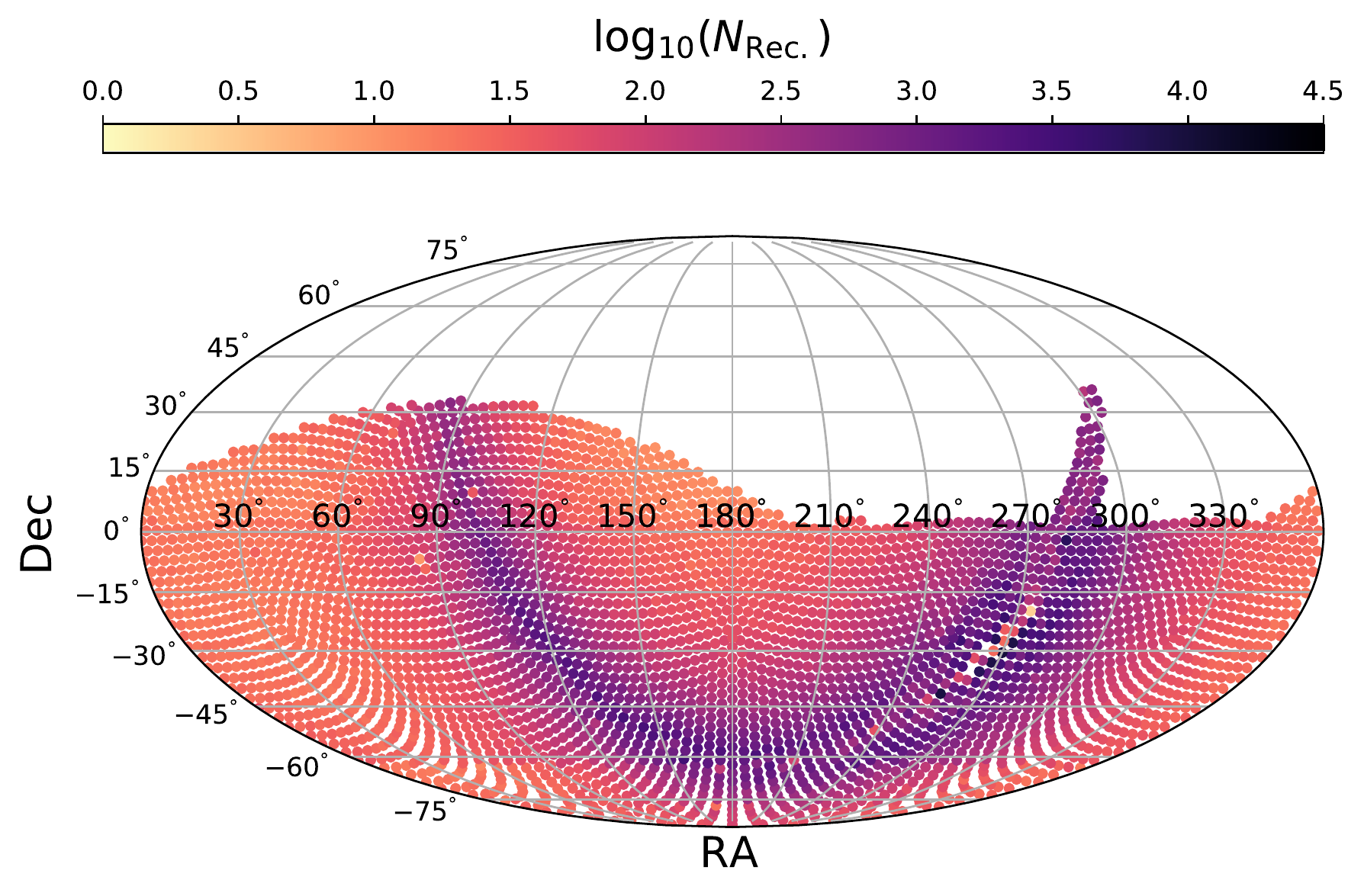}
	\plotone{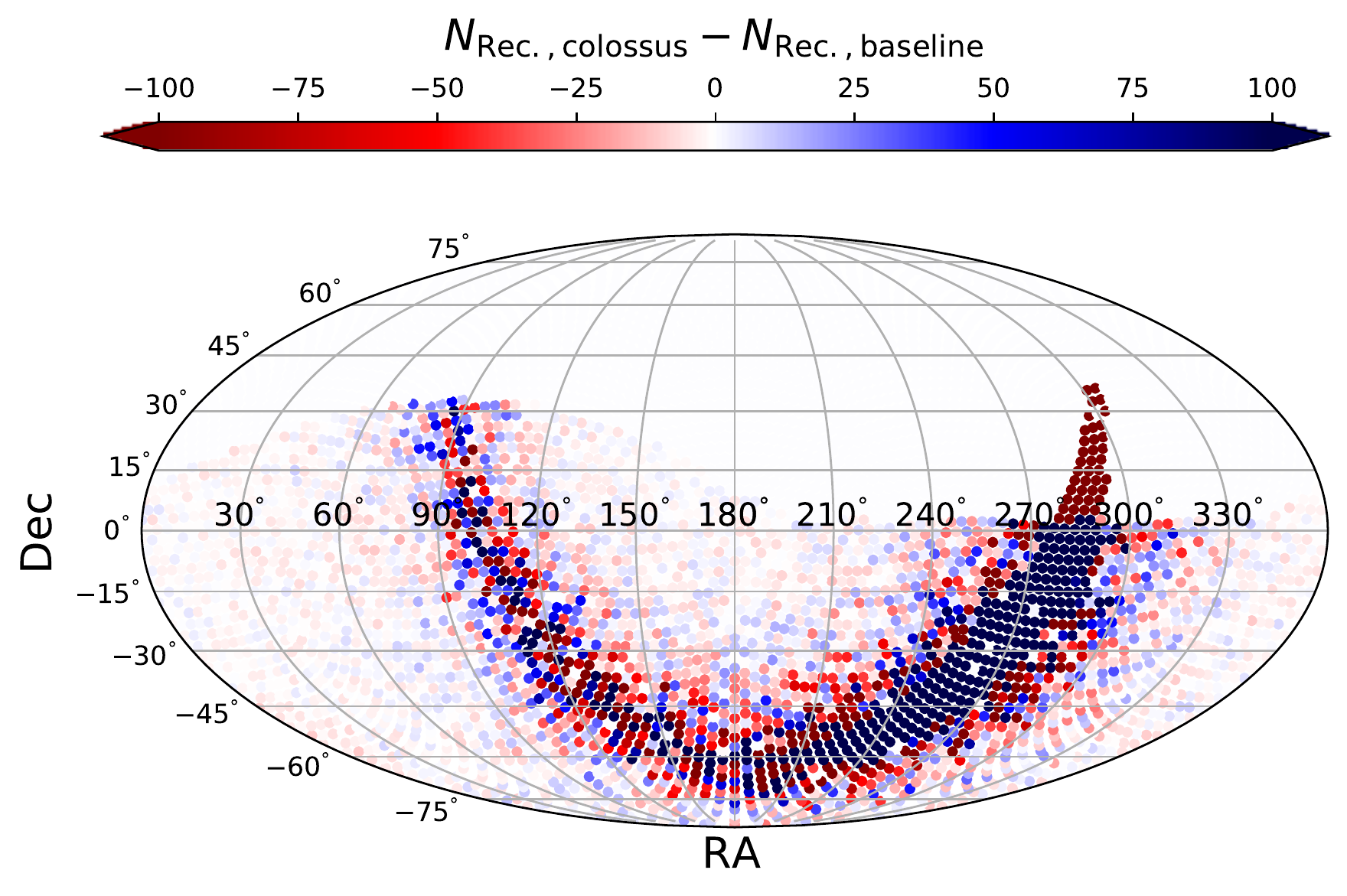}
	\caption{The number of \textit{recoverable} binaries ($N_\mathrm{Rec.}$) predicted by our galactic field simulations.  We show the results for \baseline\ cadence in the top panel.  For fields with zero \textit{recoverable} binaries, we do not plot a point.  In the bottom panel we show the difference in $N_\mathrm{Rec.}$ between the \baseline\ and \colossus\  cadences.   \label{f:recNmoll}}
\end{figure}

\begin{figure}[!t]
	\plotone{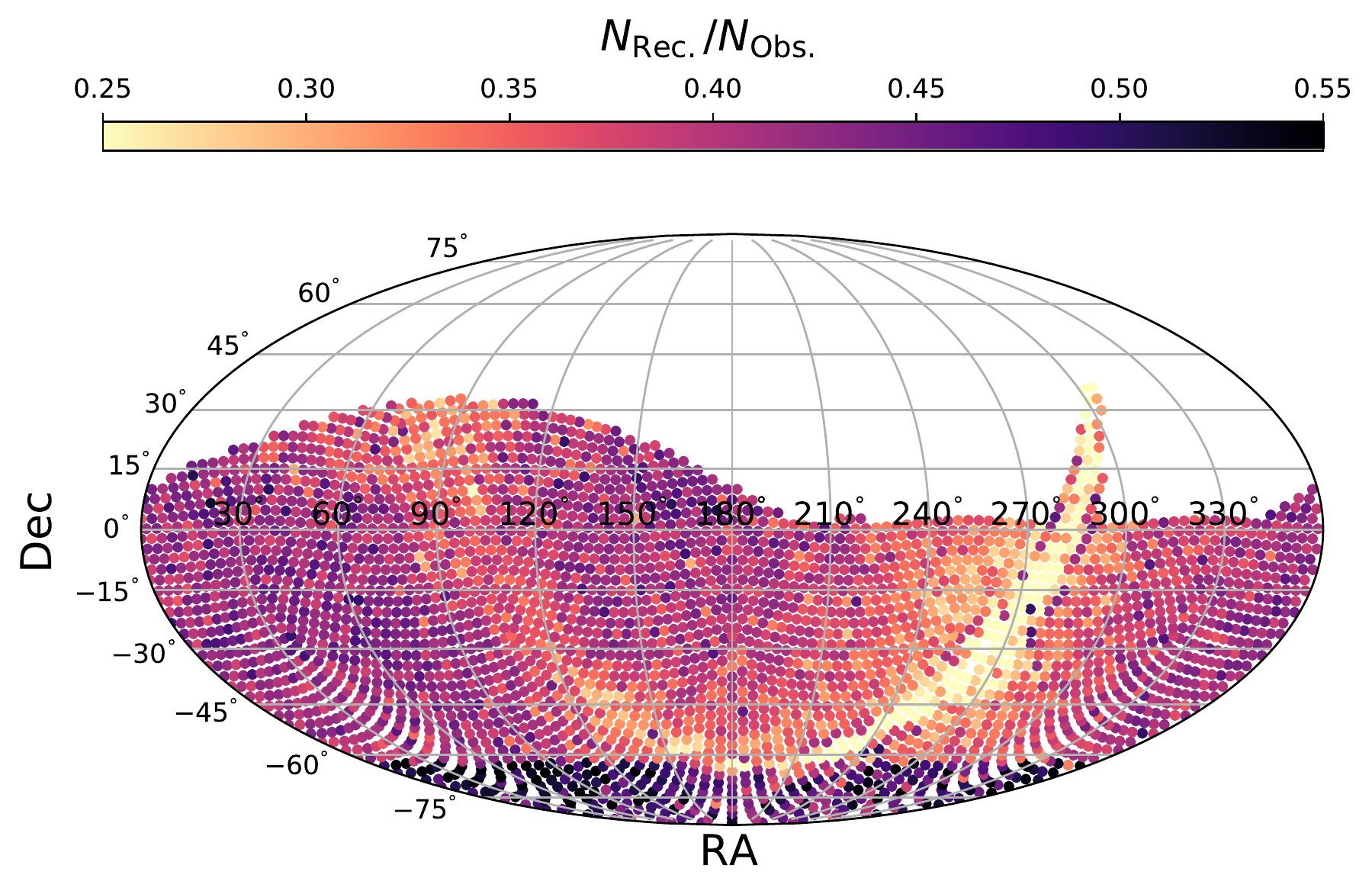}
	\plotone{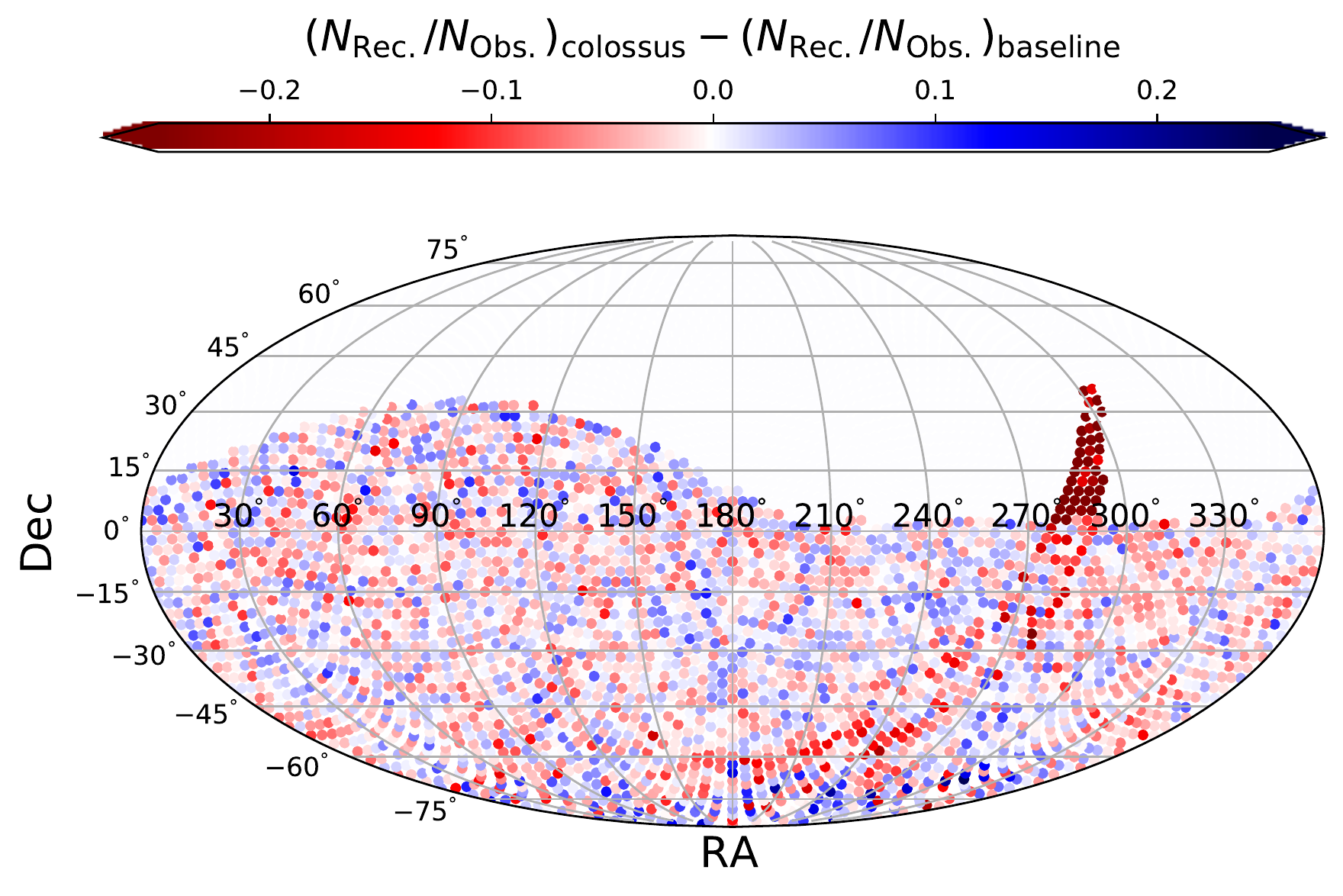}
	\caption{The ratio of the number of \textit{recoverable} binaries over the number of \textit{observable} binaries ($N_\mathrm{Rec.}/N_\mathrm{Obs.}$) as predicted by our galactic field simulations.  We show the results from the \baseline\ cadence in the top panel.  In the bottom panel we show the difference in this ratio between the \baseline\ and  \colossus\ cadences.  \label{f:pctmoll}}
\end{figure}

\vspace{2em}
\subsection{Galactic Field}

EBs from the galactic field dominate our total numbers (as expected).  In the field, our models predict that using the \colossus\ cadence would add $\sim6\times10^5$ more \textit{recoverable} detached EBs as compared to the \baseline\ cadence.  It is important to also note that the rate of recovery (that is the fraction of the number of \textit{recoverable} EBs divided by the number of \textit{observable} EBs) is lower for the \colossus\ sample.  This is primarily due to crowding in the fields within the galactic plane, and also due to a smaller number of observations on average in the other fields outside the plane.  Nonetheless, the drop in recovery rate is not nearly sufficient to erase the large increase in raw number. 

In Figures~\ref{f:recHists}-\ref{f:pctmoll} we investigate the predicted distributions of recovered EBs in each Rubin Observatory observing field, comparing the \baseline\ cadence to the \colossus\ cadence.  Beginning with Figure~\ref{f:recHists}, we show the distributions of recovery statistics for all fields in each cadence; overall the distributions are similar, though with minor difference.  Most importantly, the \colossus\ cadence produces more fields with very large numbers of \textit{recoverable} binaries (left panel); these fields come from near to the galactic center, where the raw number of stars is also highest.  The two center panels in Figure~\ref{f:recHists} show that the \colossus\ cadence produces fewer low-recovery and low-observation fields, respectively, both of which are due to the omission of the northern part of the galactic disk in the \colossus\ cadence. Interestingly, the third panel from the left shows an apparent bimodal distribution; this is due to the increased crowding and reddening in fields near to the galactic plane (that have a low frequency of observable stars) as compared to fields far from the galactic plane (with a higher frequency of observable stars). Finally, the rightmost plot in Figure~\ref{f:recHists} shows that over most fields, the recovery efficiency is similar between the two cadences.

Figure~\ref{f:recNmoll} shows the number of \textit{recoverable} binaries in each observing field for each cadence as viewed on the sky. One important feature here lies in the galactic plane and toward the galactic center (roughly at RA of 266$^\circ$ and Dec.\ of $-29^\circ$), where the \colossus\ cadence recovers a far larger number of EBs.  Figure~\ref{f:pctmoll} shows the fraction of the number of \textit{recoverable} EBs divided by the number of \textit{observable} EBs in our model.  Recall that the overall recovery fraction is larger for the \baseline\ cadence (Table~\ref{t:summary}). Figure~\ref{f:pctmoll} shows that the recovery fraction in the galactic plane is similar for both cadences (though a bit smaller in the \colossus\ cadence); the large increase in the number of \textit{observable} EBs in the \colossus\ cadence produces the large increase in the number of \textit{recoverable} EBs. 

\begin{figure*}[!t]
    \epsscale{1.15}
	\plotone{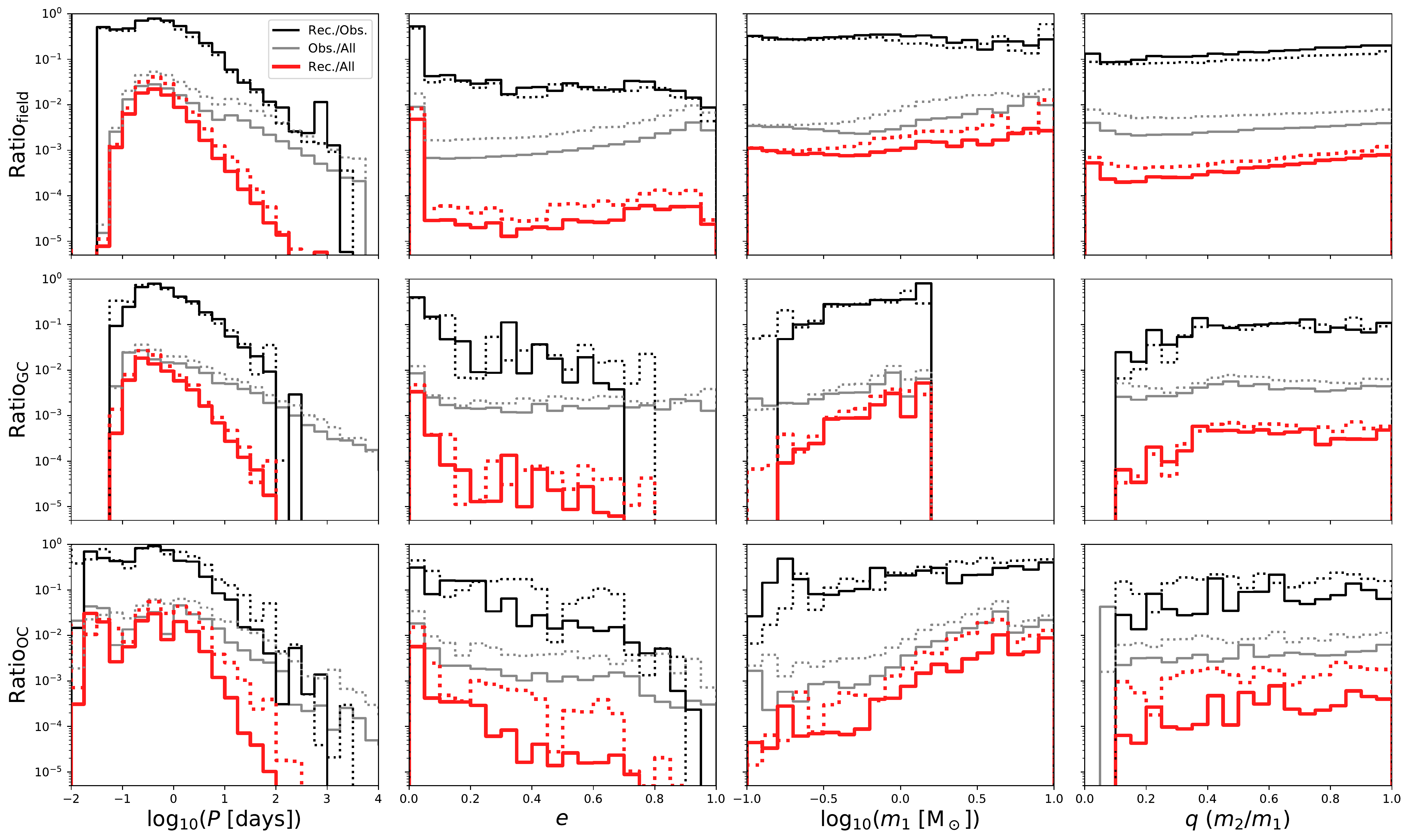}
	\caption{Recovery and observation ratios for the galactic field (top), GC (middle) and OC (bottom) samples, following a similar format to Figure~\ref{f:recdists}.  Here, each line shows a different ratio  comparing the full sample (``All"),  \textit{observable} sample (``Obs."), and \textit{recoverable} sample (``Rec."), as defined by the legend in the upper left panel.  \label{f:recratios}}
\end{figure*}

In Figures~\ref{f:recdists}~and~\ref{f:recratios}, we marginalize over the sky location in order to show one-dimensional histograms of the recovery statistics for binary periods, eccentricities, primary masses and mass ratios over the full Rubin Observatory survey area.  Here, we will focus on the top panel which shows results for the galactic field sample. Beginning with the period distribution, as expected, shorter period binaries are generally easiest to recover (and note again that we have artificially truncated the input period distribution at long periods, and we account for that truncation statistically where necessary).  We notice a decrease in our recovery efficiency toward the shortest period binaries when comparing against the ``All" sample (red and gray lines); this is primarily due to our exclusion of non-detached binaries in our \textit{observable} sample.  Moving to the eccentricities, circular binaries are far easier to detect than non-circular binaries.  A circular EB will have an eclipse of the primary and secondary at equal intervals along the orbit. However, eccentric EBs are not guaranteed to have symmetric eclipse geometries, and do not occur at equally spaced intervals.  Recovering orbital periods for eccentric binaries is also a known limitation of the Lomb-Scargle method, especially when not accounting for additional harmonics.  The overall recovery efficiency is not very sensitive to the binary mass or mass ratio.  However, if we limit our sample to investigate only binaries in our models that contain giants, we see that the recovery efficiency drops off precipitously at mass ratios $\lesssim$0.4, due to the large difference in luminosity between giants and main-sequence stars, especially at large mass ratios.

Also in Figures~\ref{f:recdists}~and~\ref{f:recratios}, we compare the distributions for the \baseline\ (solid lines) and \colossus\ (dotted lines) cadences.  Overall, the distributions resulting from these two observing cadences follow each other closely, though with an offset in total number.  The \colossus\ cadence does appear to recover more binaries with higher primary masses, likely due to the larger number of binaries coming from fields near the galactic center.

The ratios shown in Figure~\ref{f:recratios} can also be understood as potential incompleteness corrections; if we can construct the observed distributions for the real EBs recovered by the Rubin Observatory, we could use these curves to recover the intrinsic distributions (within limits).  This will be of particular interest for studies of the galactic binary population as a whole.

\subsection{Star Clusters}

Investigating the summary statistics in Table~\ref{t:summary} for the star clusters, shows once again that the \colossus\ cadence would enable more EBs to be \textit{observable} and \textit{recoverable}.  The GCs show similar recovery rates comparing the \baseline\ and  \colossus\ models.  On the other hand, the OCs show a remarkable increase in the recovery rate, from 23.8\%  $\pm$ 2.3\% using \baseline\ up to 32.2\% $\pm$ 2.3\% using \colossus\ (a $>3\sigma$ increase).  This is due to the increased observing frequency for the \colossus\ cadence within the galactic plane, where most OCs reside,.  The total number of \textit{recoverable} binaries could increase by a factor of $\sim$1.6 in GCs and a factor of $\sim$3.0 in OCs if the Rubin Observatory used the \colossus\ cadence.

The bottom two rows of Figures~\ref{f:recdists}~and~\ref{f:recratios} show the distributions of periods, eccentricities, masses and mass ratios for the GCs and OCs.  Many of the same results discussed above for the field also hold true for the clusters.  Perhaps the most interesting difference seen in these figures as compared to the field is shown in the mass panel.  The OCs will sample a much broader mass distribution (due to the generally younger age of the stars), and reach to much higher-mass stars than the GCs or the field.

\section{Discussion} \label{s:discuss}

The raw numbers of recovered binaries are dominated by the field, followed by the GCs and then OCs at much smaller raw numbers (see Table~\ref{t:summary}).  As is evident from the red lines in Figure~\ref{f:recdists}, the \colossus\ cadence is predicted to produce more EBs than the \baseline\ cadence over nearly the entire range of binary parameters studied here.  There is no clear trend in these plots comparing the differences between \baseline\ and \colossus, only an offset in raw number.  In Table~\ref{t:summary} we show that overall using a cadence that samples the galactic plane more evenly, like \colossus, may increase the total number of \textit{observable} detached EBs by a factor of $\sim$2 (adding $\sim3$ million EBs) and increase the number of \textit{recoverable} detached EBs by a factor of $\sim$1.7 (adding $\sim6\times10^5$ EBs) as compared to the standard, \baseline, cadence.

As an additional check on our model, we compare our predicted orbital period distribution of EBs with that of the EA-type EBs observed by the ASAS-SN survey \citep{kochanek2017}, in the top-left panel of Figure~\ref{f:recdists}.  The period distribution has a very similar form to that produced by our model.  At the time of writing this paper, the ASAS-SN survey has detected $\sim$48,400 EA-type EBs; our simulations suggest that the Rubin Observatory will recover nearly 20 times more EA-type EBs than ASAS-SN. Also, recall that in our simulations, we include only detached systems.  In the ASAS-SN database, only $\sim$30\% of the EBs are detached, and in the \textit{Kepler} EB database \citep{kirk2016}, the percentage of detached EBs is $\sim$40\%.  Thus if we assume we could recover all non-detached EBs, the total number of \textit{recoverable} EBs predicted by our model could increase by a factor of about 2.5 to 3.

The predicted number of \textit{recoverable} EBs in our model is in reasonable agreement with that predicted by \citet{prsa2011}; their models and analysis suggest that the Rubin Observatory could recover and characterize $\sim$6.7 million EBs with a default cadence.  This is about a factor of 6 larger than we predict in our model. There are a few possibilities for where this difference may stem from in the methods employed.  First, \citet{prsa2011} sample orbital periods from a log-uniform distribution, which weights the short-period population much more heavily than is observed (at least for binaries of roughly a solar mass), whereas we follow the empirical log-normal period distribution.  Short-period binaries are easiest to recover, and therefore including a larger sample of short-period binaries will boost recovery numbers.  As discussed in Section~\ref{ss:uncertainties}, we estimate that using a log-uniform period distribution (like \citealt{prsa2011}) would boost our number of \textit{recoverable} binaries by a factor of $\sim$3.  Second, \citet{prsa2011} extrapolate to their total number of expected binaries based on the percent of all EBs (including non-detached EBs) observed within the  \textit{Kepler} sample.  As discussed above, including non-detached EBs would increase the total number by a factor of about 2.5 to 3.  These two contributions can explain the factor of $\sim$6 difference between our overall EB recovery numbers.  We also model the galaxy in much greater detail than \citet{prsa2011}, perhaps introducing some differences in detectability (e.g., due to reddening), and we include the effects of crowding in our simulations, which can result in a significant reduction in \textit{recoverable} binaries in dense fields.  Regardless, it is safe to assume that the Rubin Observatory will be able to characterize many millions of EBs, and we should prepare for this tidal wave of data. 

Perhaps the most comprehensive list of variable stars in GCs is presented in \citet{clement2001}.  In their paper, they find about 100 EBs (including non-detached systems), and we count about 350 EBs total in their updated online catalog.   GCs are traditionally quite hard to survey photometrically from the ground due to the stellar density and distance (and photometric surveys using \textit{HST} often don't cover the full extent of the cluster), and therefore the \citet{clement2001} sample is very likely incomplete.  The Rubin Observatory will help alleviate some of these difficulties, though crowding will still be a severe limitation in GCs; on average, crowding in our model reduces the number of \textit{recoverable} EBs in GCs by $\sim$30\%.  Nonetheless, we predict that the observations from the Rubin Observatory may increase the number by known EBs in GCs by about an order of magnitude. 

OCs are generally much easier than GCs to survey for photometric variable stars from the ground.  \citet{zejda2012} presented the first catalog of variable stars in OCs by compiling available data from a set of about 535 OCs.  They found 89 detached EBs.  Though not a complete sample, this result is in rough agreement (to within a factor of a few) with the number of detached EBs we predict would be \textit{recoverable} with our methods, and therefore provides another validation check for our simulations.  We predict that roughly ten times this number of known detached EBs in OCs will be \textit{observable} by the Rubin Observatory.

We turn now to the distributions of parameters for the \textit{recoverable} binaries predicted by our model, and first we look at the black lines in Figure~\ref{f:recratios} that shows the ratio of the number of \textit{recoverable} divided by the number of \textit{observable} EBs.  In general, the distributions of recovery rates follow similar patterns for the field and cluster samples (and likewise for the \baseline\ and \colossus\ cadences).  Looking back at Figure~\ref{f:recdists}, we can see that differences in the distributions of \textit{recoverable} EBs derive from differences in the \textit{observable} distributions, which can then be mapped directly back to the intrinsic (``All") distributions.  As a more specific example, if we investigate the period distribution (left panels of Figure~\ref{f:recdists}), we see that the \textit{recoverable} EBs for OCs appear to lack short-period binaries (with periods around 1 day) as compared to the GCs and the field.  This is also true of the \textit{observable} and intrinsic period distributions for the OCs.  Physically, this is likely due to the OCs having a larger fraction of higher mass stars than the other samples; these stars have larger radii, and therefore cannot exist in orbits with these short of periods.  A similar observation can be made for the primary masses of the binaries; the \textit{recoverable} population of the OCs extends to larger masses than that of the GCs, again due to the intrinsic population in our model. 

If we compare the field to the GC and OC distributions, two-sample Kolmogorov-Smirnov (K-S) tests show that for each respective binary parameter in Figure~\ref{f:recdists}, the distributions of \textit{recoverable} binaries for each sample can be statistically distinguished from all other samples at very high confidence, except comparing the \textit{recoverable} mass-ratio distributions of the GCs and OCs (which returns a K-S $p$-value of 0.046, or $\sim$2$\sigma$).  In other words, the \textit{recoverable} EB orbital period distribution for field binaries is drawn from a different parent distribution than that of the GCs and that of the OCs (respectively).  The same statement can be made for all the parameters shown in Figure~\ref{f:recratios}, and for a comparison of these GC and OC distributions (except for the mass ratio).  

Importantly, this result indicates that observed differences between the OCs, GCs, and the field distributions derived from real observations of EBs from the Rubin Observatory, can provide important information about the intrinsic properties of the binaries.  Furthermore, simulations such as those presented here can be used to help correct for biases resulting from only observing EBs to begin to recover the intrinsic distributions of all binaries (at least within the domain accessible to the Rubin Observatory).

\subsection{Analyses Using Individual Filters}

\begin{figure}[!t]
	\plotone{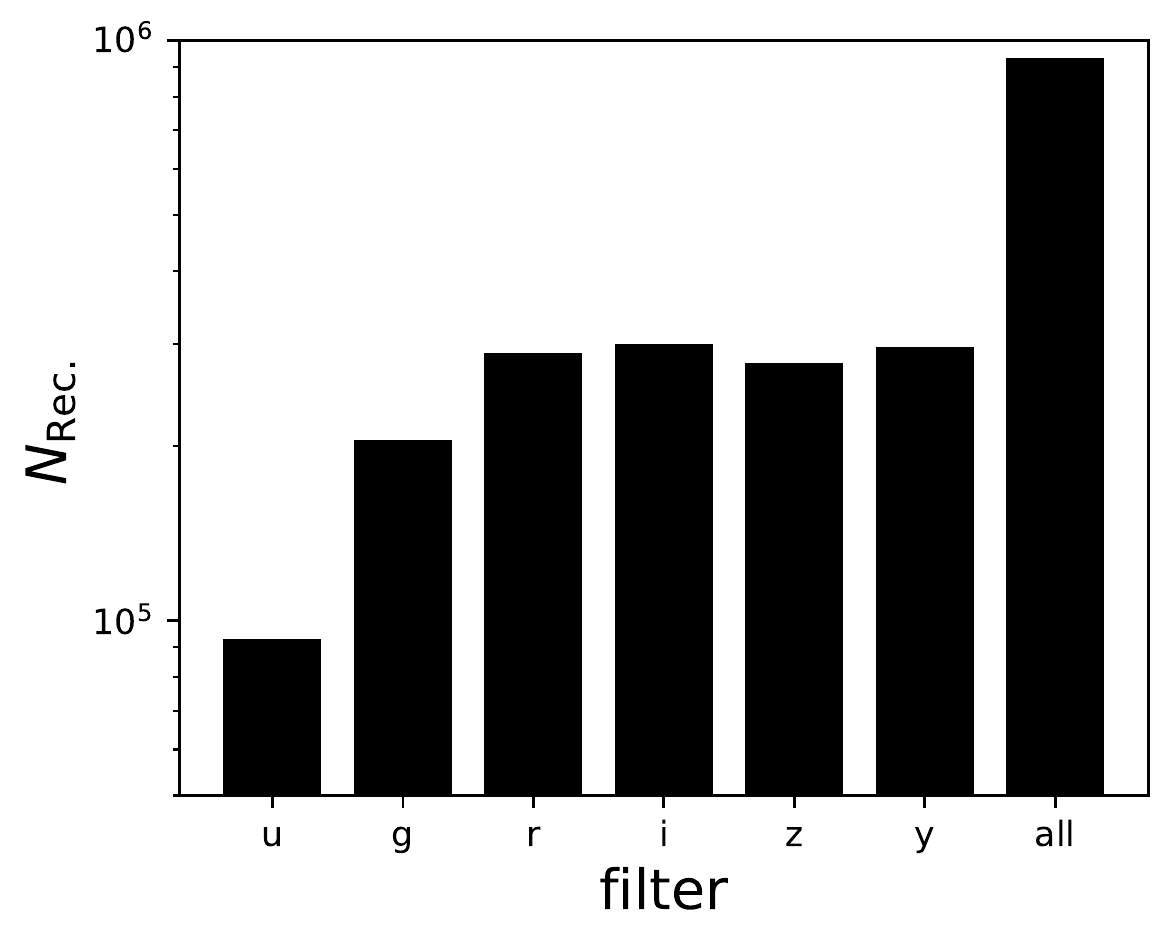}
	\caption{Number of \textit{recoverable} binaries for the \baseline\ galactic field model resulting from our single-band \texttt{gatspy} analysis using each Rubin Observatory filter ($u,g,r,i,z,y$) individually, and also using \texttt{gatspy}'s multiband method to combine all observations in all filters (``all"). \label{f:recFilt}}
\end{figure}

The results described above take all Rubin Observatory filters into account.  However we also maintain in our database results from each individual filter alone (also using \texttt{gatspy}).  We show the number of \textit{recoverable} EBs in each filter from our model for the \baseline\ galactic field simulations in Figure~\ref{f:recFilt}.  In general, including all filters in the analysis recovers $\gtrsim$3 times more binaries than any individual filter alone.  The combined information from all the additional observing times dramatically improves the ability for \texttt{gatspy} to recover the correct period for an EB.  Looking at the individual filters, the number of recovered binaries drops off as one moves blueward, due to the decreased number of observation (particularly in the $u$ band), and the fainter magnitudes from the more common lower-mass stars in the galaxy model (e.g., see Figure~\ref{f:recdists}).  Likewise, if we look at the distribution of periods of recovered binaries when only considering individual filters, the $u$ band fails to recover the longer period binaries that are reached by the redder filters.  Combining all filters allows for the widest range in periods to be recovered. 

\subsection{Double White Dwarf Binaries}

Double white-dwarf binaries (DWDs) are of particular interest, in part because they are potential SN Ia precursors, and also given a short enough orbital period the system may  be detectable in gravitational waves using the proposed Laser Interferometer Space Antenna (LISA) observatory.  White dwarfs are included in our models (coming from either \texttt{TRILEGAL} or \texttt{COSMIC}).  However, as this is not the primary interest of this paper, we do not take great care in modelling this population nor do we include white dwarf atmospheres in our simulations, and therefore the numbers and photometry are not fully accurate for these sources.  Nonetheless, it may be instructive to briefly investigate the raw numbers that our model predicts for DWDs to motivate future work.   In the galactic field, our model predicts about 100 DWDs may be \textit{observable}, with about 60\% of those being \textit{recoverable} with our current algorithm.  Inspecting the distributions of periods for the DWDs that our model considers \textit{observable} shows that they all have periods below $\sim$2 days and about 70\% have periods less than $\sim$10 hours, which may fall in the detection window of LISA. (Our models also predict many DWDs in the GCs and OCs, but none are predicted to be \textit{observable} with our current models.)

\citet{korol2017} studied the population of DWDs that are expected to result from observations by the Rubin Observatory (as well as Gaia and LISA) in great detail.  They do a far more careful job of modeling the DWD population, and use a different method for simulating the light curves and detecting the eclipses than we use here.  They predict about 1000 DWDs will be detected by the Rubin Observatory.  Given the differences in our techniques and the limited ability of our simulations to accurately model the DWD population, we are not surprised to have found a lower number of \textit{recoverable} DWDs than \citet{korol2017}.  We plan to investigate these sources further, using our simulation tool but with more accurate modelling of DWDs, in a subsequent paper.

\section{Conclusions} \label{s:conclusions}

In this paper, we present and analyze a simulation of observations of detached EBs in the galactic field, GCs and OCs expected to result from the Vera C.\ Rubin Observatory.  We take care to include a realistic population of binaries, drawing from the \texttt{TRILEGAL} galactic model for the field, and evolving binaries from realistic initial conditions using \texttt{COSMIC} for the star clusters.  We generate simulated observations of these binaries on the expected observing dates and conditions using \texttt{OpSim}, and compare two cadences: \baseline, the expected cadence (at the time of writing), and \colossus, a proposed cadence that samples the galactic plane more evenly.  We generate light curves using the \texttt{ellc} software, and then use these simulated observations as inputs to the \texttt{gatspy} software, which employs a multiband periodogram to include all observations in all the filters from the Rubin Observatory, to attempt to recover the orbital periods of the input EBs.   The use of all filters increases the number of \textit{recoverable} binaries by a factor of $\gtrsim3$ as compared to using any individual filter alone. 

Using the \baseline\ cadence, our model predicts that the Rubin Observatory will recover about one million EBs, with the vast majority coming from the galactic field.  This number is in good agreement with other estimates from the literature \citep[e.g.][when accounting for the different sample selections]{prsa2011}.  If instead the \colossus\ cadence is used, the number of detached EBs expected to be recovered increases by a factor of $\sim$1.7 in the field and GCs, and by a factor of $\sim$3 in the OCs due to their locations along the galactic plane that is sampled more heavily by the \colossus\ cadence.  Also, including non-detached binaries may boost these numbers by an additional factor of 2.5 to 3.  

We also investigate the distributions of parameters that define our sample of \textit{recoverable} binaries, and compare to those of the \textit{observable} and intrinsic (including non-eclipsing binaries) populations.  We find that statistically detectable differences should be present in the Rubin Observatory data when comparing the distributions of EB periods, eccentricities, masses and mass ratios (and other parameters) from the galactic field to those in star clusters, and also comparing the GCs to the OCs.  Our simulations suggest that these differences in the \textit{recoverable} EB populations are not a result of different recovery efficiencies in these different populations, and instead can be mapped back to the \textit{observable} and intrinsic populations.  Simulations such as these will be useful for bias-correcting the EB population recovered from the Rubin Observatory to reveal intrinsic properties of the galactic and star cluster binary populations.

\acknowledgments

We thank the anonymous referee for their very helpful comments and suggestions.  This research was supported in part through the computational resources and staff contributions provided for the Quest high performance computing facility at Northwestern University which is jointly supported by the Office of the Provost, the Office for Research, and Northwestern University Information Technology.
This material is based upon work supported by the LSST Corporation (LSSTC), through Enabling Science Grants \#2017-UG05, \#2018-UG04 and \#2019-UG01, the Northwestern Weinberg College of Arts and Sciences Undergraduate Research Grant Program, and the NASA Illinois Space Grant program.
This research used the WEBDA database, operated at the Department of Theoretical Physics and Astrophysics of the Masaryk University.  A.A.M. is funded by the Large Synoptic Survey Telescope Corporation, the Brinson Foundation, and the Moore Foundation in support of the LSSTC Data Science Fellowship Program; he also receives support as a CIERA Fellow by the CIERA Postdoctoral Fellowship Program (Center for Interdisciplinary Exploration and Research in Astrophysics, Northwestern University).
\vspace{7em}

\bibliographystyle{aasjournal}
\bibliography{references.bib}

\end{document}